\documentstyle[sprocl]{article}

\input{psfig}
\def\Journal#1#2#3#4{{#1} {\bf #2}, #3 (#4)}

\def\PLB{{\em Phys. Lett.}  B}
\def\PRL{\em Phys. Rev. Lett.}
\def\PRD{{\em Phys. Rev.} D}

\def\ra{\rightarrow}

\def\be{\begin{equation}}
\def\nn{\noindent}

\def\etal{{\it et al.}}
\def\ee{\end{equation}}
\def\bea{\begin{eqnarray}}
\def\eea{\end{eqnarray}}
\begin{document}

\rightline{\vbox{\halign{&#\hfil\cr
&SLAC-PUB-7466\cr
&May 1997\cr}}}
\vspace{0.8in}

\title{{$b \ra s\ell^+\ell^-$ IN THE LEFT-RIGHT SYMMETRIC MODEL}
\footnote{To appear in the {\it Proceedings of the $2^{nd}$ International 
Conference on B Physics and CP Violation}, Honolulu, HI, 24-27 March 1997}
}

\author{ {T.G. RIZZO}
\footnote{Work supported by the Department of Energy, 
Contract DE-AC03-76SF00515}
}

\address{Stanford Linear Accelerator Center,\\
Stanford University, Stanford, CA 94309, USA}


\maketitle\abstracts{We begin to analyze and contrast the predictions for the 
decay $b\ra s\ell^+\ell^-$ in the Left-Right Symmetric Model(LRM) with those of 
the Standard Model(SM). In particular, we show that the forward-backward 
asymmetry of the lepton spectrum can be used to distinguish the SM from the 
simplest manifestation of the LRM.}

\section{Introduction}

The study of rare $B$ decays may provide us with a window into new physics 
beyond the SM. In particular, the decays $b\ra s \gamma$~{\cite {joa}} and 
$b\ra s\ell^+\ell^-$~{\cite {joa2}} may arguably provide the cleanest 
environment for such searches since they are both reasonably well understood 
within the SM and most of the difficulties associated with hadrodynamics are 
avoided. In the LRM, the decay $b\ra s \gamma$ has already been examined and 
many interesting features were uncovered~{\cite {old}}. In particular it was 
shown that left-right mixing terms can be enhanced by a helicity 
flip factor of $\sim m_t/m_b$. Here we turn to the decay 
$b\ra s\ell^+\ell^-$~{\cite {vol}}. In order to analyse this mode we 
use the following procedure which is now relatively standard: ($i$) Determine 
the complete operator basis 
for the effective Hamiltonian, ${\cal H}_{eff}$, responsible for $b\ra s$ 
transitions in the LRM; ($ii$) evaluate the coefficients of these operators at 
the weak scale; ($iii$) run these coefficients down to the relevant low energy 
scale $\mu \sim m_b$ via the RGE's and take the appropriate matrix 
elements; ($iv$) calculate observables. We outline these four steps in what 
follows with the details to be found elsewhere~{\cite {big}}.

The decay rate for $b\ra s\ell^+\ell^-$ , including QCD 
corrections, is computed via the following effective Hamiltonian, 
\begin{equation}
{\cal H}_{eff} = {4G_F\over\sqrt 2}\sum_{i=1}^{12}C_{iL}(\mu)
{\cal O}_{iL}(\mu)+L\ra R \,,
\end{equation}
which is evolved from the electroweak scale down to $\mu\sim m_b$ by the 
RGE's.  The ${\cal O}_{iL,R}$ are the set of operators involving only the light 
fields which govern $b\to s$ transitions.  The complete basis for each helicity 
structure consists of the usual six 4-quark operators ${\cal O}_{1-6L,R}$, 
the penguin-induced electro- and chromo-magnetic operators respectively 
denoted as ${\cal O}_{7,8L,R}$, as well as 
${\cal O}_{9L,R}\sim e\bar s_{L,R\alpha}\gamma_\mu b_{L,R\alpha}\bar\ell
\gamma^\mu\ell$, and ${\cal O}_{10L,R}\sim e\bar s_{L,R\alpha}\gamma_\mu 
b_{L,R\alpha}\bar\ell\gamma^\mu \gamma_5\ell$ which arise from box diagrams 
and electroweak(EW) penguins. In the LRM we not only have the 
augmentation of the operator basis via the obvious doubling of $L\ra R$, 
but two new additional 4-quark 
operators of each helicity structure are also present at the tree-level due to 
a possible mixing between the $W_{L,R}$ gauge bosons:  
${\cal O}_{11L,R}\sim (\bar s_\alpha \gamma_\mu c_\beta)_{R,L}
(\bar c_\beta \gamma^\mu b_\alpha)_{L,R}$ and ${\cal O}_{12L,R}\sim 
(\bar s_\alpha \gamma_\mu c_\alpha)_{R,L}(\bar c_\beta \gamma^\mu 
b_\beta)_{L,R}$. Note that the extension of the operator basis implies that 
the conventional model-independent analysis of $b\ra s\gamma$ and 
$b\ra s\ell^+\ell^-$ by Hewett~{\cite {joa2}} will not 
apply in this case.

\section{Analysis}

The determination the matching conditions for the 24 operators at the 
EW scale is 
somewhat cumbersome since the LRM contains a very large number of free 
parameters and, in addition to new tree graphs, 116, one-loop graphs must also 
be calculated.(Additional diagrams due to possible physical Higgs exchange are 
not yet included.) For simplicity, we will assume that the 
$Z-Z'$ mixing angle is zero, the $W-W'$ mixing angle($\phi$) is real, 
right-handed neutrinos  
are heavy($m_N \gg m_b$) and that the $Z'$ and $W'$ masses are correlated 
through 
the usual relationship that follows from $SU(2)_R$ breaking via Higgs 
triplets~{\cite {tgr}}. All remaining parameters, in particular the 
right-handed version of the CKM matrix, $V_R$, are left arbitrary. Using the 
results in Refs.~{\cite {joa2,old}}, the RGE analysis is relatively 
straightforward with the $24\times 24$ anomalous dimension matrices breaking 
into two $12\times 12$ identical sets as the $``L"$ and $``R"$ operators are 
decoupled and do not mix under RGE evolution. This RGE running is performed 
at essentially full NLL.

For $b\ra s \ell^+\ell^-$, the effective Hamiltonian above leads to the matrix 
element (neglecting the strange quark mass) 
\begin{eqnarray}
{\cal M} & = &  {\sqrt 2 G_F\alpha\over\pi}\Bigg[ C_{9L}^{eff}
\bar s_L\gamma_\mu b_L\bar\ell\gamma^\mu\ell+C_{10L}\bar s_L\gamma_\mu
b_L\bar\ell\gamma^\mu\gamma_5\ell \nonumber \\
& & \quad\quad  -2C_{7L}^{eff} m_b\bar s_L i\sigma_{\mu\nu}{q^\nu\over q^2}
b_R\bar\ell\gamma^\mu\ell+L\ra R\Bigg] \,,
\end{eqnarray}
where $q^2$ is the momentum transferred to the lepton pair. Note that 
$C_{9L,R}^{eff}$ contains the usual phenomenological long distance 
terms and that all the CKM elements are now contained in the 
coefficients themselves. From here we can directly 
obtain the expression for the double differential decay distribution 
\begin{eqnarray}
{d\Gamma\over {dz ds}} & \sim & {3\over {4}}\beta (1-s)^2 \Bigg\{ \left[
(a_L^2+a_R^2)+(b_L^2+b_R^2)\right]{1\over {2}}\left[(1+s)-(1-s)\beta^2 z^2
\right] \nonumber \\
& & \quad\quad \left[(a_L^2-a_R^2)-(b_L^2-b_R^2)\right]\beta zs +4x(a_La_R+
b_Lb_R) \nonumber \\
& & \quad\quad +{4\over {s^2}}(C_{7L}^2+C_{7R}^2)(1-s)^2(1-\beta^2 z^2) \\
& & \quad\quad -{2\over {s}}Re\left[C_{7L}(a_L+a_R)+C_{7R}(b_L+b_R)\right]
(1-s)(1-\beta^2 z^2) \Bigg\} \,, \nonumber
\end{eqnarray}
where $z=\cos \theta_{\ell \ell}$, $s=q^2/m_b^2$, $x=m_\ell^2/m_b^2$, 
$\beta=\sqrt {1-4x/s}$, $a_{R,L}=C_{9L}^{eff}\pm C_{10L}+2C_{7L}/s$ and 
$b_{L,R}=a_{L,R}(L\ra R)$. We normalize this rate to the usual semileptonic 
branching fraction($B=0.1023$), including finite $m_c/m_b=0.29$ and QCD 
corrections with $\alpha_s(M_Z)=0.118$. LRM corrections to the semileptonic 
rate are, of course, also included; here the assumption that $m_N>m_b$ 
becomes relevant.

\section{First Results}

Since the LRM parameter space is so large, we have only begun to probe its 
intricacies. Let us look here at a rather simple 
example where $V_L=V_R$ and the $SU(2)_{L,R}$ gauge couplings are equal; this 
is the so-called ``manifest" LRM. In 
this case the $K_L-K_S$ mass difference and direct Tevatron collider searches 
require{~\cite {soni}} that $W_R$ be heavy; we take 
$M_{W_R}=1.6$ TeV so that $t_\phi=\tan \phi$ is now the only free parameter 
since $W_R$ contributions are now almost completely decoupled. 
Fig.1 shows the prediction for the $b\ra s\gamma$ branching fraction in this 
case and we see that the SM result is essentially obtained when $t_\phi=0$, 
apart from a very small correction of order $M_{W_L}^2/M_{W_R}^2$, but 
also that a conspiratorial solution occurs 
when $t_\phi \simeq -0.02$. The results of the CLEO experiment{~\cite {cleo}} 
are also shown. From the $b \ra s\gamma$ perspective these two cases are 
indistinguishable, 
independent of what further improvements can be made in the branching 
fraction measurement. 

\vspace*{-0.5cm}
\nn
\begin{figure}[htbp]
\centerline{
\psfig{figure=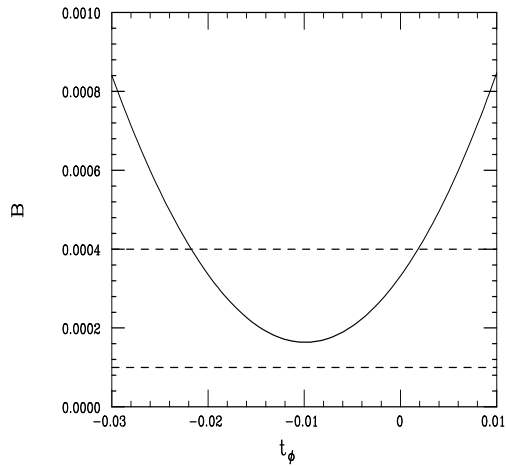,height=7.5cm,width=7.5cm,angle=-90}}
\vspace*{-0.9cm}
\caption{Prediction for the $b\ra s\gamma$ branching fraction for 
$m_t(m_t)=170$ GeV as a function of the tangent of the $W-W'$ mixing angle in 
the LRM at NLL for the case discussed in the text. The $95\%$ CL CLEO results 
lie inside the dashed lines.}
\end{figure}
\vspace*{0.4mm}

Can $b\ra s\ell^+\ell^-$ be used to distinguish these two cases? Fig.2 shows 
both the differential invariant mass distribution of the lepton pair as well 
as the forward-backward asymmetry for these two scenarios. ($m_N=300$ GeV was 
assumed here but the results are found to be 
insensitive to this choice.) While it is clear 
that the two decay distributions are very similar and cannot separate the two 
scenarios, it is obvious that the predictions for the 
asymmetry are quite different particularly in the highly sensitive region 
below the $J/\psi$ peak. It is in this region that one has the most 
sensitivity to interference between the terms involving one of the $C_{7L,R}$ 
operators and terms proportional to $C_{9,10L,R}$. In fact a $\chi^2$ fit to 
Monte Carlo data generated 
with the LRM as input is very poor if we allow for the existence of only the 
SM operators. Other observables, such as the polarization of final state 
$\tau$'s lack this sensitivity. 
This simple demonstration shows the added power of the 
observables associated with the $b\ra s\ell^+\ell^-$ decay and their ability to 
distinguish models with new physics from the SM. 
The analysis presented here only scratches the surface of the LRM giving us 
a flavor for what is possible; a more detailed study of the possible structure 
of $V_R$ will be given elsewhere{\cite {big}}.

\vspace*{-0.5cm}
\nn
\begin{figure}[htbp]
\centerline{
\psfig{figure=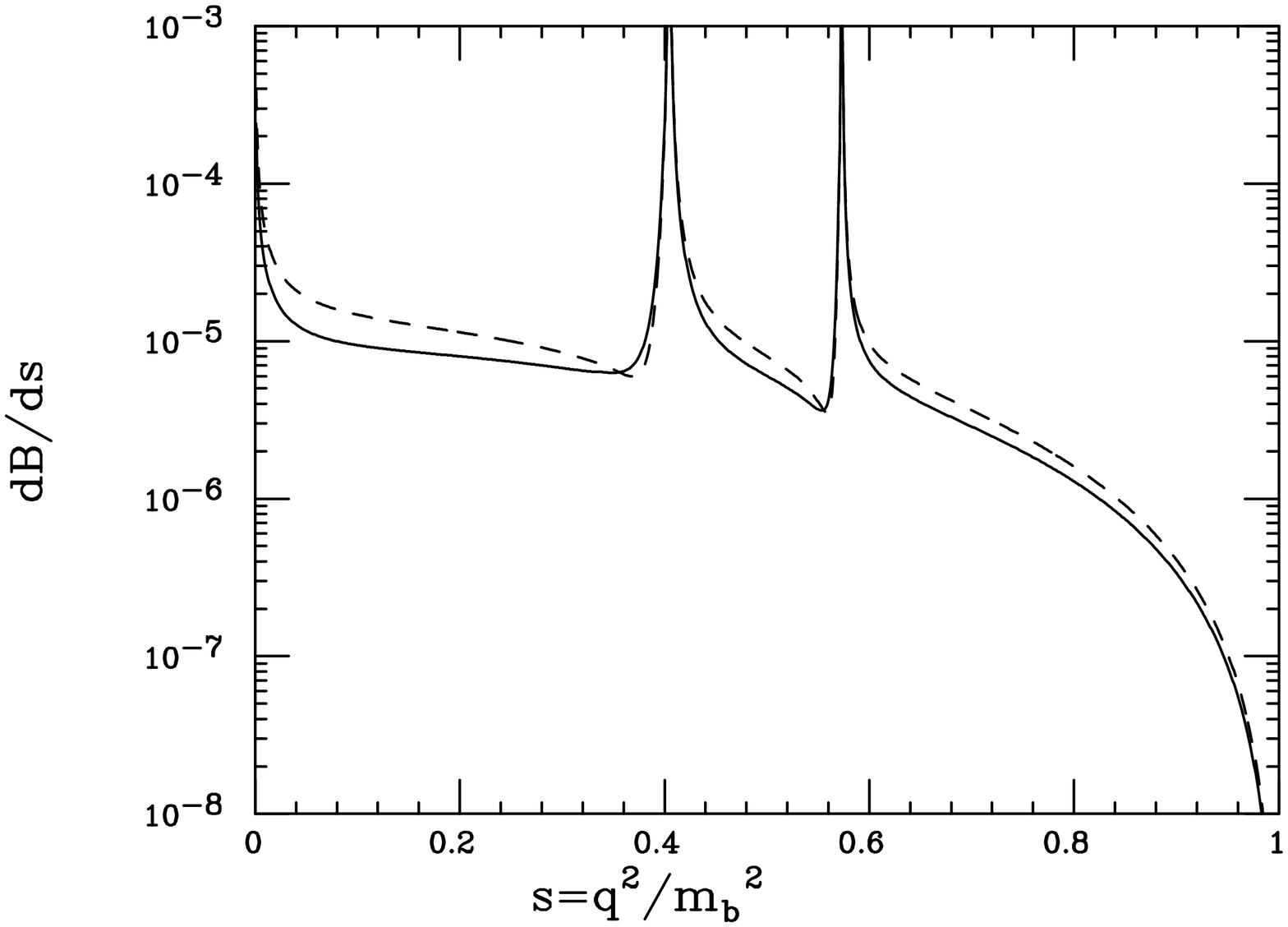,height=6.4cm,width=6.4cm,angle=-90}
\hspace*{-5mm}
\psfig{figure=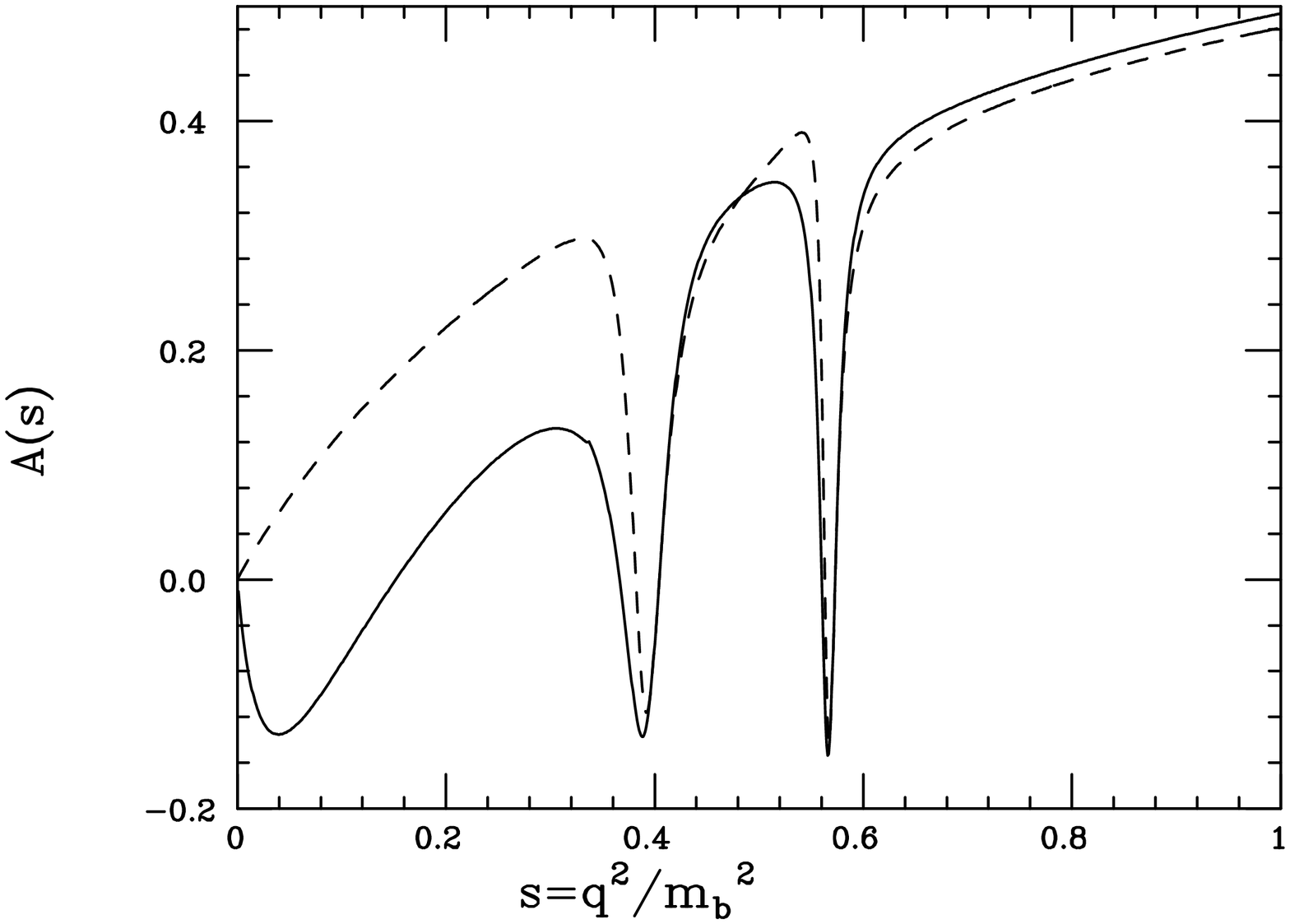,height=6.4cm,width=6.4cm,angle=-90}}
\vspace*{-0.6cm}
\caption{Differential decay distribution and lepton forward-backward 
asymmetry for the decay 
$b\ra s\ell^+\ell^-$ in the SM(solid) and LRM(dashed) for the case discussed 
in the text. The lepton mass is ignored.} 
\end{figure}
\vspace*{0.1mm}

\section*{Acknowledgments}

The author would like to thank J. Hewett, A. Masiero, G. Isidori and A. Kagan 
for discussions related to this work.

\end{document}